%% file: main.tex
\documentclass[sigplan,nonacm]{acmart}
\usepackage{xspace,multirow}
\usepackage{paralist}
\usepackage{graphicx}
\newcommand{\sysname}{Spice\xspace}

\iffalse
\newcommand{\ben}[1]{\textcolor{red}{Ben: #1}}
\newcommand{\josh}[1]{\textcolor{blue}{Josh: #1}}
\newcommand{\bsd}[1]{\textcolor{orange}{Baltasar #1}}

\else
\newcommand{\ben}[1]{}
\newcommand{\josh}[1]{}
\newcommand{\bsd}[1]{}
\fi

\newcommand{\nextdraft}[1]{}

\begin{document}
\title{Taming Serverless Cold Starts Through OS Co-Design}
\fancyhead{} 
\author{Ben Holmes, Baltasar Dinis, Lana Honcharuk, Joshua Fried, Adam Belay}
\affiliation{
    \country{}
  \institution{MIT CSAIL}}

\begin{abstract}
\input{sections/abstract.tex}
\end{abstract}

\maketitle

\input{sections/intro.tex}

\input{sections/motiv.tex}

\input{sections/approach}

\input{sections/detailed_design.tex}
\input{sections/impl.tex}

\input{sections/eval.tex}

\input{sections/related}

\input{sections/discussion}
\input{sections/conclusion}

\bibliographystyle{ACM-Reference-Format}
\bibliography{ref}

\end{document}

%% file: sections/abstract.tex
Serverless computing promises fine-grained elasticity and operational simplicity, fueling widespread interest from both industry and academia. Yet this promise is undercut by the cold start problem, where invoking a function after a period of inactivity triggers costly initialization before any work can begin. Even with today’s high-speed storage, the prevailing view is that achieving sub-millisecond cold starts requires keeping state resident in memory.

This paper challenges that assumption. Our analysis of existing snapshot/restore mechanisms shows that OS-level limitations, not storage speed, are the real barrier to ultra-fast restores from disk. These limitations force current systems to either restore state piecemeal in a costly manner or capture too much state, leading to longer restore times and unpredictable performance. Furthermore, current memory primitives exposed by the OS make it difficult to reliably fetch data into memory and avoid costly runtime page faults.

To overcome these barriers, we present \sysname, an execution engine purpose-built for serverless snapshot/restore. \sysname integrates directly with the OS to restore kernel state without costly replay and introduces dedicated primitives for restoring memory mappings efficiently and reliably. As a result, \sysname delivers near-warm performance on cold restores from disk, reducing latency by up to 14.9$\times$ over state-of-the-art process-based systems and 10.6$\times$ over VM-based systems. This proves that high performance and memory elasticity no longer need to be a trade-off in serverless computing.

%% file: sections/intro.tex
\section{Introduction}
\label{sec:introduction}

\begin{figure}
    \centering
    \includegraphics[width=\linewidth]{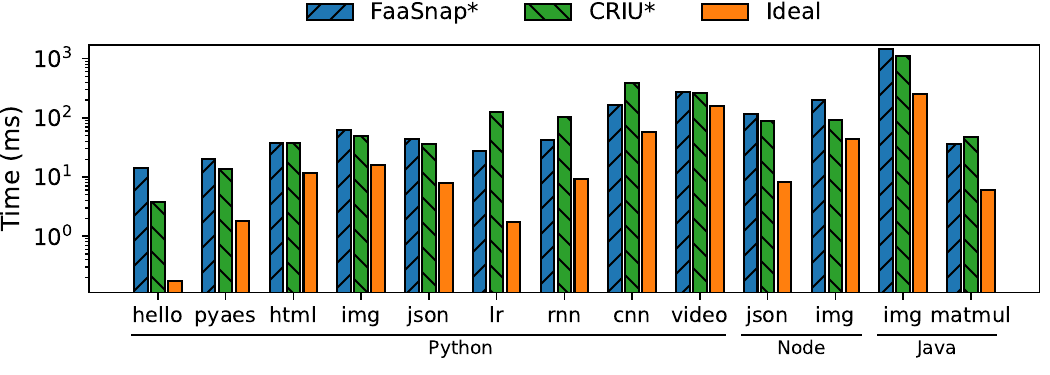}
    \caption{The performance gap in snapshot/restore: Even tuned, state-of-the-art systems remain orders of magnitude slower than the ideal restore time. Note the logarithmic y-axis.}

    \label{fig:ideal}
\end{figure}

Serverless computing promises effortless elasticity: developers deploy lightweight functions, while the platform dynamically provisions resources in response to demand~\cite{oneforwardtwoback}. Freed from managing infrastructure, users benefit from fine-grained billing and scalability, while providers can, in theory, achieve high utilization by scheduling workloads just-in-time. In today’s commercial platforms (e.g., AWS Lambda~\cite{awslambda}, Azure Functions~\cite{azurefunctions}), these functions are typically packaged in containers or lightweight VMs, which must be initialized before running user code. However, this vision is fundamentally limited by cold starts: the latency incurred when a function is invoked on a machine with no prior cached state~\cite{reap,catalyzer,faasnap,mitosis,seuss,faasm,xfaas}.

Cold start delays stem from multiple sources, including container setup, language runtime initialization, library loading, and function-specific startup logic such as JIT compilation~\cite{replayable-exec,pronghorn,fireworks}. These steps often add tens to hundreds of milliseconds --- frequently longer than the function’s actual execution~\cite{afaas} --- making them a major source of user-visible latency. Collectively, these overheads undermine the responsiveness of serverless platforms.

Existing mitigation strategies fall into two broad categories. \emph{Warm-state} approaches, such as keeping containers alive for a short window~\cite{firecracker,faascache}, deliver low-latency startup but consume memory proportional to the number of functions retained. Recent techniques like fork-based replication~\cite{mitosis,cxlfork,trenv,afaas} reduce this cost by enabling replication of a single warm instance across cores or machines. These systems (e.g., leveraging fork locally or using remote memory mechanisms such as RDMA or CXL) allow multiple invocations to share memory copy-on-write, reducing total memory usage while preserving low startup latency. In essence, they amortize one “parent” process across many children, but still require at least one copy of each function to remain alive in memory. This model, however, fundamentally assumes a warm instance is available somewhere on the cluster. For the many functions in the long tail of an invocation distribution, or during a cold start after a period of inactivity, this assumption often does not hold, leaving no fast path for the initial request.

On the other hand, \emph{cold-state} approaches don't rely on warm state present~\cite{reap,faasnap,sabre,prebaking}. These approaches often serialize initialized function state to persistent storage and reload it on demand, offering a natural fit for serverless platforms where functions may be invoked rarely but must start quickly. In principle, these snapshot/restore approaches can achieve both elasticity and memory efficiency. In practice, however, restore latencies remain far higher than warm starts, preventing widespread adoption. 

To quantify this gap, Figure~\ref{fig:ideal}  compares two state-of-the-art systems against a pessimistic \emph{ideal} baseline across several representative serverless workloads (detailed in Table~\ref{tab:workloads}). We use an asterisk(*) to denote that we modified both Faasnap (VM-based) and CRIU (process-based) for optimal performance. For Faasnap, we selected the best-performing of its prefetching strategies, and for CRIU, we eliminated several known overheads. Our ideal baseline is calculated as the sum of the time to read the snapshot’s working set from storage and the function's warm execution time, pessimistically assuming no overlap between I/O and computation. Even against these tuned systems and this pessimistic baseline, a significant performance gap remains.

This failure is not due to hardware limits, but to a deeper mismatch between serverless requirements and the interfaces provided by today’s OSes. Current OS abstractions are optimized for incremental startup, not for the bulk restoration of an already-initialized process. This mismatch forces existing systems to rely on slow, general-purpose mechanisms. One of the clearest examples is metadata restoration. Lacking dedicated kernel support, process-level tools like CRIU~\cite{criu} must replay the original setup through a long sequence of expensive system calls. The alternative --- snapshotting an entire virtual machine~\cite{reap,faasnap,sabre} --- avoids this replay but captures far more state than necessary, introducing new overheads including scheduling interference from the guest OS. Crucially, both approaches are bottlenecked by a second major challenge: restoring memory contents efficiently.

Memory restoration faces two intertwined challenges. The first is how to proactively populate a process’s memory with its working set (the subset of pages needed for execution). One option is to synchronously load the entire predicted working set before execution. While this guarantees the pages are resident in memory, it can result in long delays before execution begins. The alternative is to prefetch asynchronously, overlapping I/O with the function's startup routine. This promises lower latency in theory, but is unreliable with today's OS interfaces and doesn't eliminate stalls due to faults.
Second, an application’s memory is a composite of shared, file-backed pages and private, modified data. Current OS interfaces cannot restore this complex structure efficiently, forcing a choice between slow, page-by-page updates or sacrificing memory sharing altogether. As fast local storage becomes the norm, it is this two-fold interface mismatch --- not raw I/O bandwidth --- that has become the dominant factor in cold start latency.

To resolve this interface mismatch and eliminate these OS-level overheads, we present \sysname, a snapshot/restore system co-designed with a new set of OS primitives for fast memory and metadata restoration. By providing these missing mechanisms, \sysname makes restoring from persistent storage fundamentally practical. Our evaluation shows that \sysname reduces the overhead from restoring from a cold disk to under 5ms for a range of functions of varying complexity across several language runtimes. 

This paper makes the following contributions:
\begin{itemize}
    \item A study of existing snapshot/restore techniques that identifies persistent overheads due to mismatches between the requirements of low-latency restore and today's OS interfaces.
    \item A novel OS metadata restoration process that bypasses costly syscall replay by deserializing process state in a single, batched operation from a co-designed snapshot format.
    \item New kernel mechanisms for memory management that enable fast, reliable restore of process memory that eliminates page fault overhead during startup.
    \item The implementation and evaluation of \sysname, a prototype engine demonstrating that our approach can reduce cold restore latency to under 5ms, substantially outperforming state-of-the-art systems.
\end{itemize}

Our findings reframe the conventional trade-offs in serverless design, demonstrating that snapshot/restore is a first-class primitive capable of delivering both low latency and high memory efficiency for serverless systems.

%% file: sections/motiv.tex
\section{Background and Motivation }
\label{sec:background}

This section analyzes the performance of the two predominant snapshot/restore strategies introduced in Section~\ref{sec:introduction}: process-level restore, exemplified by CRIU~\cite{criu}, and VM-based snapshots, used by systems like REAP~\cite{reap}, FaaSnap~\cite{faasnap}, and Sabre~\cite{sabre}. We demonstrate how both approaches are fundamentally constrained by the lack of direct OS support for rapid restoration, leading to the two core bottlenecks foreshadowed earlier: reinstating process metadata and repopulating memory. Our analysis focuses on these core restoration costs; we do not measure the overhead of initializing container primitives like cgroups and network namespaces, as techniques for accelerating these are orthogonal and well-explored~\cite{rund, trenv, sigmaos, flashcube}.

\subsection{The Challenge of State Reconstruction}
\label{sec:forced-approaches}

\begin{figure}
    \centering
    \includegraphics[width=\linewidth]{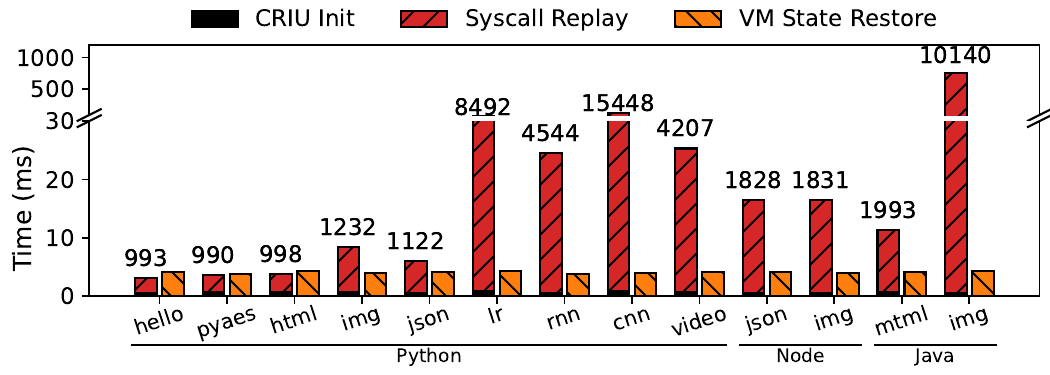}
    \caption{Initialization overheads for VM- and process-based restore; total syscall counts for replay are shown above the bars. Process restore costs grow with process complexity due to syscall replay. VM restore costs are flat but still non-trivial, reflecting fixed hypervisor operations.} 
    \label{fig:init-overheads}
\end{figure}

\begin{figure}
    \centering
    \includegraphics[width=\linewidth]{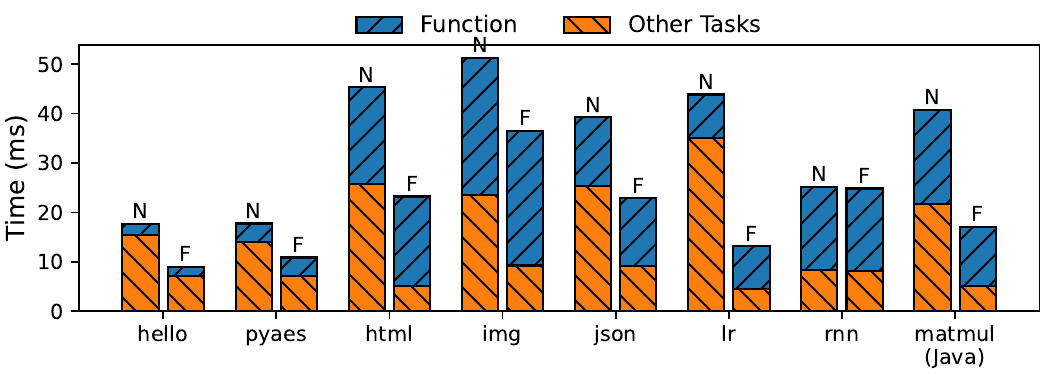}
    \caption{Scheduling delay after VM restore with the function running in \texttt{SCHED\_NORMAL} (N) and \texttt{SCHED\_FIFO} (F). Even with \texttt{SCHED\_FIFO}, the function competes with kernel threads and daemons restored from the guest image.}
    \label{fig:sched-delay}
\end{figure}

The fundamental challenge for fast restoration is that modern operating systems lack a native interface to restore a process's kernel-managed state. This state includes a wide array of resources—threads, file descriptors, memory mappings, signal handlers, and more. Not only is there no single command to reinstate this entire collection at once, but even restoring a single resource to its previous condition can require a complex sequence of operations.

This lack of an efficient restore primitive forces systems into one of two extreme approaches: either (1) replaying the system calls that originally created the process, or (2) snapshotting and restoring the entire virtual machine it runs within. Both strategies push existing abstractions beyond their intended uses and are responsible for significant cold-start overheads.

\textbf{Syscall replay.} 
At one extreme, restore begins from an empty process and attempts to reconstruct every piece of kernel state --- threads, file descriptors, memory mappings, and more --- by reissuing the original system calls that created them. This strategy is expensive: there are often hundreds or thousands of such calls required, as most kernel objects lack dedicated import or restore interfaces to directly set them to their previous state.
For example, restoring a single file descriptor may require a sequence of syscalls: \texttt{open} to create the file, \texttt{lseek} to set its offset, \texttt{dup} to assign the correct descriptor number, and \texttt{fcntl} to reapply flags.
{Figure~\ref{fig:init-overheads}} quantifies this overhead, showing that the latency of process-based restore grows with application complexity, from a simple Python function to a more complex JVM application, as the number of syscalls increases from hundreds to thousands. Each resource adds to the latency, and since few can be reconstructed lazily, all must be completed before execution resumes. 

The orchestration of this process from userspace adds further overhead. Tools like CRIU stage snapshot data in temporary memory regions at restore time, inject a restorer binary into otherwise unused space, transfer control to that binary, unmap its own code, and finally remap the snapshot data into the correct virtual addresses. This elaborate dance avoids even more costly options such as using a second process to manipulate the target with \texttt{ptrace}.

\textbf{Full-VM snapshots.}
At the other extreme, systems restore entire virtual machines from snapshots. This approach bypasses syscall replay entirely, as all kernel data structures remain intact. As shown in {Figure~\ref{fig:init-overheads}}, restore operations are reduced to a small, fixed set of hypervisor actions, such as reinitializing vCPU state, reattaching devices, and restoring other host-managed state. These complete in just a few milliseconds regardless of guest complexity.

A major downside, however, is that reviving the whole VM also revives the entire guest kernel and all its runtime activity --- not just the target function. After a prolonged pause, the guest must immediately resume deferred kernel housekeeping and background services. This includes periodic tasks, daemons, and maintenance threads, all of which are scheduled as soon as the VM becomes active.
{Figure~\ref{fig:sched-delay}} illustrates this problem, showing that a function can be delayed by several milliseconds due to contention with guest kernel tasks like RCU reclamation. We found that even raising the function’s scheduling priority with \texttt{SCHED\_FIFO} offers only partial relief, as critical kernel threads continue to interrupt the function's execution. This scheduling interference significantly impacts end-to-end latency.

\medskip
\noindent
\emph{Takeaway:} In the absence of kernel support for reinstating process state directly, systems are pushed into two extremes. 
Replay-based restore suffers from syscall and orchestration overheads, while VM snapshots avoid replay but revive the entire guest kernel, whose deferred housekeeping and background services cause scheduling interference that heavily impacts function startup latency. 
Neither approach achieves the millisecond-scale responsiveness required for serverless workloads.

\subsection{The Challenge of Rapid Memory Restoration}
\label{sec:memory-restoration}

While kernel metadata must be reinstated eagerly, memory contents can be restored lazily. This is attractive in serverless environments where functions often touch only a small portion of their memory. Accordingly, existing systems~\cite{reap,faasnap,sabre} typically rely on demand paging, bringing pages into memory only when accessed. However, naive demand paging introduces significant latency, as every missing page triggers a blocking fault. To mitigate this, systems employ prefetching to load the predicted working set into memory before it is accessed.

\begin{figure}
    \centering
    \includegraphics[width=\linewidth]{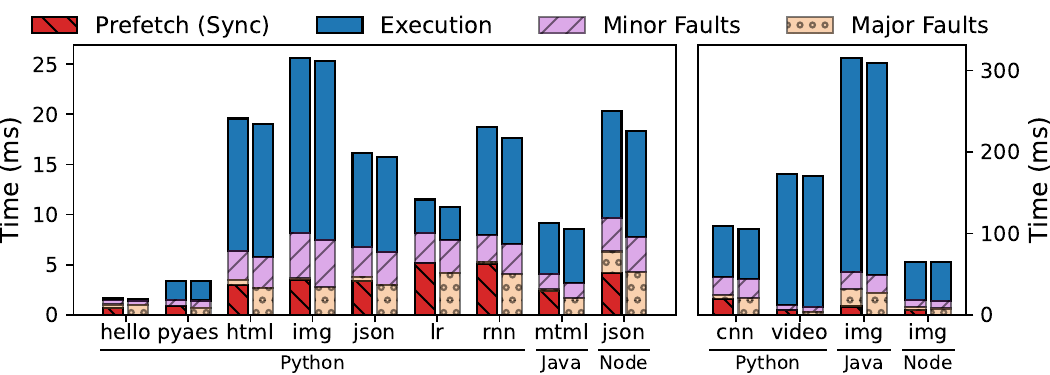}
    \caption{Impact of prefetching technique on execution latency. Synchronous prefetching with \texttt{read} (left bars) delays execution, while overlapping prefetching with execution using \texttt{madvise} (right bars) fails to properly ensure that pages are actually fetched. Neither approach resolves costly minor faults. 
    }
    \label{fig:prefetch-importance}
\end{figure}

However, as {Figure~\ref{fig:prefetch-importance}} shows, current prefetching strategies are insufficient. REAP~\cite{reap} employs synchronous prefetching, which loads all pages before execution and minimizes major page faults but results in a long, fixed delay upfront. FaaSnap~\cite{faasnap} improves upon REAP's working set estimation and also adds asynchronous prefetching, which attempts to overlap I/O with execution.
Unfortunately, this strategy does not meaningfully improve end-to-end latencies, and neither strategy resolves the slowdown incurred by minor faults, caused by missing PTEs or writes to copy-on-write (CoW) pages.

\textbf{No Interface for Guaranteed Population.}
The fundamental challenge in pre-warming data is the operating system's lack of a reliable, non-blocking interface to populate its page cache. While a synchronous \texttt{read()} call can guarantee data is fetched from disk, it does so by stalling the calling thread until the I/O is complete, which results in delayed execution. Consequently, systems must rely on asynchronous, advisory mechanisms through interfaces \texttt{madvise()}. However, this is merely a hint, not a command; the kernel retains full discretion to ignore the request, act on it partially, or de-prioritize it based on internal heuristics like memory pressure. This unpredictability means the main application thread remains vulnerable to major page faults on data that was requested but never loaded, ultimately forcing any prefetching strategy into a best-effort approach with no performance guarantees.

\textbf{The Lingering Cost of Page Faults.}  
Even when a page has been successfully fetched into memory, it is not usable until the kernel installs a page table entry (PTE) mapping the virtual address to its physical location. This happens lazily and incurs a \emph{minor fault} on first access, requiring a kernel trap to update the page table. For large heaps or runtime data structures, thousands of such faults can accumulate quickly.
Worse, if a fetched page is written to, it will incur an additional fault and copy. File-backed pages are mapped copy-on-write (CoW) to preserve shared page cache state. On the first write, the kernel must allocate a private copy and copy the original contents—an expensive operation. In real-world measurements, JVM-based functions incur tens of thousands of CoW faults during startup, contributing tens of milliseconds of delay.

In virtualized environments, these fault costs become even more severe. Each fault—minor or CoW—forces a VM exit to the hypervisor, incurring hundreds of cycles of latency before fault handling can even begin. Even moderate levels of fault activity can translate into long tail latencies for restored functions running inside VMs like Firecracker.

\begin{figure}
    \centering
    \includegraphics[width=\linewidth]{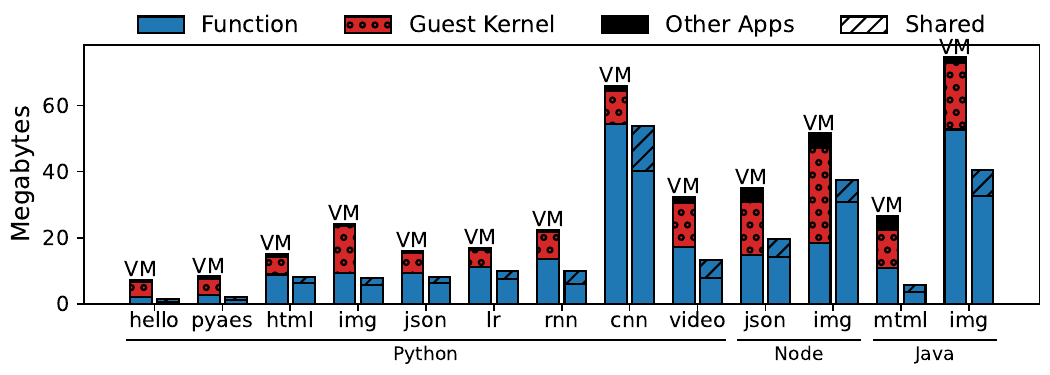}
    \caption{Working set composition for VM-based and process-based restore. VMs include substantial kernel-mapped memory and guest-level state, while processes benefit from the potential of sharing 17-51\% of working set pages through the host page cache with other functions.}
    \label{fig:ws-composition}
\end{figure}

\textbf{Barriers to Efficient Memory Sharing}
While prefetching optimizations are important, an orthogonal and equally powerful strategy is to reduce the number of pages that must be fetched in the first place. A natural opportunity arises from reusing memory pages that are already resident in the OS page cache, such as those belonging to shared libraries (e.g., libc) or language runtimes. Figure~\ref{fig:ws-composition} shows that such shared pages can constitute up to 50\% of a function’s working set; thus, reusing them could, in principle, halve the volume of data that must be restored from a snapshot.

Existing OS interfaces, however, provide no efficient mechanism for realizing this idea. A process’s memory regions, known in Linux as Virtual Memory Areas (VMAs), often contain a mixture of unmodified pages (identical to their backing file) and modified, private pages (unique to the process). The OS tracks this distinction only at the fine granularity of individual page table entries (PTEs), not at the coarser VMA level, and it exposes no system call that allows mapping a file while selectively overlaying a small set of private pages.

As a result, restore systems typically fall back on inefficient workarounds. One approach, as implemented in CRIU, is to map the file and then traverse the private pages, issuing a series of costly, one-by-one system calls to update the corresponding PTEs. Another is to fragment what should be a single contiguous memory region into many smaller VMAs in order to isolate modified pages, which inflates kernel metadata and increases memory overhead, and adds to restore latency~\cite{replayable-exec}. Crucially, both strategies require work for every modified range of pages across the application’s memory, even for regions that are unlikely to be accessed again, making it impossible to apply overlays lazily or restrict them to the actual working set.

In the context of VM snapshots, the situation is even more restrictive: shared memory reuse across functions is infeasible because the hypervisor treats guest memory and disk as opaque blocks of memory. This opacity prevents the host from identifying and reusing file-backed pages that may already exist in the host’s page cache.

\medskip
\noindent
\emph{Takeaway:} Prefetching helps avoid reinstating unused memory, but today’s kernel interfaces make it difficult to do so efficiently. Taking advantage of cached pages is cumbersome, asynchronous prefetching is unreliable, and minor and CoW faults --- even after prefetch --- add tens of milliseconds of delay, especially under virtualization. Meaningful improvement will require new kernel mechanisms for precise, non-blocking memory restoration.

\subsection{Fork is Not a Panacea}
\label{sec:fork-systems}

Recent work has explored fork-based approaches~\cite{mitosis,cxlfork,trenv,afaas}, which achieve near-zero startup latency by cloning a warm process. Within a single machine, these systems leverage the kernel’s existing \texttt{fork()} mechanism to replicate process state efficiently. This operation is fast because the kernel can duplicate internal data structures in place and relies on copy-on-write to defer actual page copies. In this model, metadata restoration is essentially free: file descriptors, mappings, and page tables are inherited wholesale from the parent process rather than reconstructed.

These fork-based approaches are highly effective at rapid, horizontal scaling from a warm parent. Their design, however, makes a distinction between cloning a running instance and restoring that first parent from a cold state. The engineering focus of these systems, particularly for remote fork, is on optimizing high-speed memory replication and cross-node transport. The problem of efficiently instantiating the initial parent from persistent storage, therefore, lies outside their primary optimization path.

This creates a crucial trade-off: forks provide rapid scaling as long as a parent is available, but they do not address the cold-start latency of creating that parent. When all instances have exited, these systems must fall back on slower, conventional methods for instantiation. Our work addresses this specific phase, focusing on an efficient cold restore that can precede the first fork.

\subsection{Summary of Kernel Limitations}
Our analysis reveals that existing restore strategies, whether process- or VM-based, are fundamentally constrained by structural limitations in modern operating systems. The core bottlenecks are twofold: process-based systems are hampered by slow metadata reconstruction, requiring thousands of expensive system calls to rebuild in-kernel state, while both approaches suffer from inefficient memory restoration, relying on unreliable prefetching and a fault-driven process to populate page tables. These issues are compounded by platform-specific overheads, such as scheduling interference within VMs and inefficient memory sharing in container setups. Even fork-based approaches, which excel at cloning warm instances, must ultimately rely on these slower methods when a function is not warm in the cluster.

Ultimately, all current designs expose the same underlying gap: today’s OS interfaces are built for incremental startup, not the rapid restoration of a complete process state. Closing this gap requires new kernel mechanisms that make restore itself a first-class, efficient operation. In the next section, we introduce \sysname, which addresses these structural problems directly.

%% file: sections/approach.tex
\section{Approach}
\label{sec:approach}

\begin{figure}
    \centering
    \includegraphics[width=\linewidth]{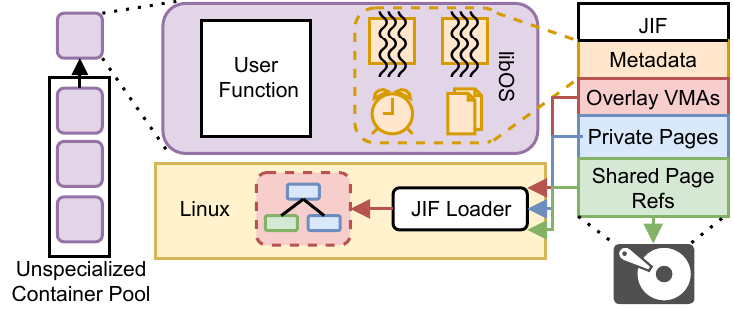}
    \caption{\sysname Overview}
    \label{fig:placeholder}
\end{figure}

Based on the limitations of existing systems, we designed \sysname, a new system that leverages novel {operating system mechanisms} for low-latency function restoration. 
Our design is driven by the goal of minimizing end-to-end latency while maximizing resource utilization and flexibility. To achieve this balance, our approach is guided by three design goals commonly considered when optimizing cold starts:

\begin{compactenum}
\item Minimize the critical-path state by lazily deferring any non-essential initialization.

\item Amortize generic setup costs by performing all function-agnostic preparation ahead of time.

\item Maximize parallelism by overlapping I/O operations with computational work during restoration.
\end{compactenum}

In line with these principles, we choose to operate on process-level snapshots within a container rather than full VMs, as the container is the standard unit of serverless deployment. In addition to reducing the amount of state required to fetch at invocation time (Figure~\ref{fig:ws-composition}), this also provides flexibility: our approach is effective both in bare-metal deployments and within virtual machines, since the function is ultimately executed as a containerized process in either environment. To amortize setup costs, our design utilizes a {host-side pool of pre-initialized, unspecialized containers} that are ready to be specialized, eliminating container setup costs from the critical path. Our model assumes fast storage is available to load the snapshot into one of these containers quickly.

To realize these ideas, \sysname introduces new OS mechanisms to address the core challenges of slow metadata reconstruction and inefficient memory restoration. Effectively addressing these issues requires first-class OS support, as existing interfaces are designed for process creation, not restoration. As a full implementation would demand deep changes to the Linux kernel, we built a prototype in {Junction}~\cite{junction}, a container system that that implements the Linux kernel interface in userspace (similar to gVisor). Its architecture is ideal for rapid prototyping and offers full control over the address space layout, which helps avoid the memory mapping conflicts that challenge systems like CRIU. Our ultimate goal is to provide a blueprint for these mechanisms, hoping our results will encourage the Linux kernel community to adopt first-class support for high-performance process restoration. Our prototype design allows us to implement new metadata interfaces entirely in userspace, while our memory management improvements are built as a module in the host kernel. Figure~\ref{fig:placeholder} shows our system's design, which we now describe in more detail.

\subsection{The JIF File Format}
To orchestrate the restore process, our approach relies on a new snapshot file format co-designed with our system's architecture. Inspired by the Executable and Linkable Format (ELF) that tells an operating system how to load a program, we introduce the Joint Image Format (JIF). The JIF is a structured binary format that packages all the information needed to restore a process --- its serialized kernel metadata, memory pages, and layout information --- into a single, self-contained file. This unified format is explicitly designed to be parsed efficiently, enabling \sysname to handle metadata and memory restoration in parallel.

\subsection{Efficient Metadata Restoration}
\label{sec:metadata-approach}

The metadata section of the JIF is generated by a set of intelligent, per-subsystem serialization mechanisms. Instead of treating kernel state as an opaque blob, these interfaces leverage semantic knowledge of each subsystem to create a maximally compact representation. For example, the networking and IPC serializers automatically trim empty pipe and socket buffers, ensuring that only essential in-flight data is saved in the snapshot.

This highly optimized format is key to eliminating the overhead of syscall replay. \sysname restores state directly from this compact binary representation, reconstructing the process without replaying individual system calls. Because Junction is a userspace library operating system (libOS), the entire deserialization of this state happens within the Junction runtime, requiring no system calls or kernel transitions. This has two key benefits:
\begin{compactenum}
    \item {No Kernel-Entry Overhead:} By handling the entire metadata restore as a single, batched operation within userspace, we avoid the cost of thousands of individual kernel crossings inherent to the syscall replay model.
    \item {Lazy Resolution and I/O Overlap:} The userspace restore process can be fully overlapped with the I/O required to read the snapshot. Where possible, we defer expensive operations. For example, file descriptors are not fully re-opened on restore; instead, they are resolved lazily on their first use, avoiding costly file system traversals during the critical startup path.
\end{compactenum}

\subsection{High-Performance Memory Management}
\label{sec:memory-approach}

While Junction manages most of the application's kernel state, the host kernel still manages page tables and memory mappings. \sysname introduces a dedicated memory management system in the host kernel with two primary areas of improvement.

\subsubsection{Optimized Prefetching}
To solve the problems of unreliable, hint-based prefetching and fault-driven page table population, we introduce a new kernel prefetching module. This module leverages the layout of the JIF, which identifies the predicted working set and stores all its constituent pages in a contiguous block. This design is critical for performance, as it enables the module to reliably fetch the entire working set with a single, high-throughput I/O operation, avoiding the overhead of many small reads. After fetching the data, the module pre-populates the corresponding page table entries (PTEs), eliminating the thousands of minor faults that plague existing systems.

The module treats memory differently based on its expected usage to avoid expensive copy-on-write (CoW) faults:
\begin{itemize}
    \item Pages from the snapshot that were modified during initialization but are {not written to} during typical execution are mapped as Copy-on-Write, allowing them to be shared safely.
    \item Pages that are known to be private and written to during restore to are installed directly into writable memory, bypassing the CoW mechanism entirely.
\end{itemize}
Leveraging the host's page cache provides a powerful mechanism for accelerating frequently invoked functions. By loading pages that are mapped copy-on-write—including both shared libraries and private, read-mostly data—into the cache, the OS can retain this information even after a container is torn down. This enables subsequent invocations to launch rapidly from the warm cache, avoiding the significant resource overhead of explicitly keeping idle container instances alive. This caching benefit does not apply to private, frequently-written pages, whose contents are unique to a single invocation. \sysname is designed to handle this distinction, dynamically choosing whether to leverage the page cache for reusable data or restore private memory directly for writable pages based on memory usage patterns.

\subsubsection{Sharing with Overlay VMAs}
To maximize memory sharing, we introduce a new kernel structure called an {Overlay VMA} (Figure~\ref{fig:overlay_vma}). This allows \sysname to efficiently restore a memory region that is mostly shared but contains a sparse set of private, modified pages. The Overlay VMA maps the shared backing file and uses a compact B-tree, stored in the JIF, as an auxiliary data structure. This B-tree serves as a sparse representation of the original page table, indicating which pages are private, modified, or simply zero-filled; this encoding allows us to completely avoid storing or fetching zero pages from the snapshot.

When a page fault occurs within an Overlay VMA, the kernel consults this B-tree to determine how to handle the fault. The tree dictates whether the faulting address should be served from the private overlay in the snapshot, the shared backing file, or by mapping a new zero page. While our optimized prefetching aims to prevent page faults entirely for ideal, profiled workloads, this B-tree mechanism is essential for guaranteeing correctness in case execution diverges. This approach provides a complete solution, avoiding address space fragmentation and costly page-by-page updates without sacrificing correctness.



%% file: sections/detailed_design.tex
\section{Mechanisms for Memory Restoration}
\label{sec:detailed-design}
\begin{figure}[t]
    \centering
    \includegraphics[width=\linewidth]{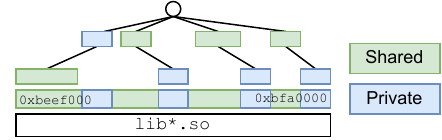}
    \caption{An Overlay VMA}
    \label{fig:overlay_vma}
\end{figure}

\begin{figure}[t]
    \centering
    \includegraphics[width=\linewidth]{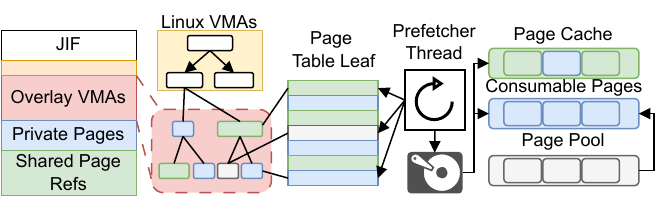}
    \caption{\sysname Detailed Memory Restore Design.}
    \label{fig:kernel_module}
\end{figure}

Figure~\ref{fig:kernel_module} details the components and operation of \sysname's memory module, which is designed to faithfully recreate the memory subsystem's state as efficiently as possible. Achieving this requires deep integration with memory management to meet its primary goals: prefetching the precise working set to restore memory contents quickly, installing page table entries upfront to avoid page faults, and efficiently preserving memory sharing.

\subsection{JIF preparation}

In an offline phase, functions are pre-warmed with multiple invocations to trigger operations like Just-In-Time (JIT) compilation before a full-memory checkpoint is taken.
Memory pages from file-backed mappings are compared with their backing in storage; unmodified pages are discarded from the snapshot, leaving only a reference to the original file.
Finally, Overlay VMAs are generated for each original VMA to efficiently represent the overlay of private pages on shared mappings.
The Overlay VMA trees are pre-balanced and stored in a compact binary format that requires no de-serialization at restore time.

After the initial checkpoint is taken, the system performs several invocations to compute an ordered trace of the function's memory accesses. The trace is added to the JIF file, and the involved pages are relocated and placed in a contiguous range in their order of access. This layout is explicitly optimized to enable high-throughput sequential reading during restoration.

\subsection{Restore Process}

To minimize latency, the memory module is designed to overlap VMA creation with prefetching. These two critical tasks begin the moment a restore is initiated.

Instead of creating and inserting VMAs individually, \sysname performs a single batch allocation. This process is accelerated because the pre-computed B-trees for Overlay VMAs are stored in the JIF in a ready-to-use binary format. This allows these complex structures to be directly slotted into the kernel’s memory management framework, taking expensive VMA manipulation off the critical path.

Simultaneously, prefetching begins with an initial synchronous I/O batch for the small set of pages the process will access first. This allows execution to resume immediately while the bulk of the memory is fetched by a dedicated prefetcher thread. The prefetcher thread works continuously, interleaving new I/O requests with the installation of PTEs for the previous batch. Eagerly installing PTEs eliminates thousands of minor page faults and ensures pages are ready for the application just before they are needed, preventing execution stalls.

This high-throughput approach is supported by a zero page pool, which is essential for taking page allocation off the critical path. Submitting large batches of I/O requires acquiring many physical pages quickly, which can exhaust fast per-core free lists and fall back to the slower global allocator. The zero page pool provides a ready supply of pre-allocated pages for two purposes: buffering incoming data from the snapshot and satisfying requests for writable zero pages. This prevents the prefetcher from stalling and allows it to keep pace with the application’s memory demands.

%% file: sections/impl.tex
\section{Implementation}
\label{sec:impl}
\input{sections/tab-workloads}

We implement \sysname's metadata snapshot and restore interface within Junction's library OS~\cite{junction}. Junction provides a containerized environment for running unmodified Linux binaries, offering strong isolation by handling most system calls in userspace within a restrictive \texttt{seccomp} and \texttt{chroot} jail. While we do not use its kernel-bypass features, Junction's userspace management of kernel state is ideal for our purposes. It simplifies metadata serialization and deserialization, allowing us to demonstrate the benefits of a dedicated restore interface. We use the lightweight \texttt{cereal} library to generate our compact metadata representation. To aid in manipulating, updating, and verifying JIF images, we built \texttt{jiftool} in 7,400 lines of Rust.

The core memory restoration logic resides in a 4,300-line Linux 6.5.0 kernel module. This module also handles working set estimation. We found that performing this tracing in the kernel is critical for accuracy; the high overhead of userspace tracing tools can stall execution and distort memory access patterns. To generate a stable and accurate profile, our process involves iteratively tracing multiple restores, using the working set from the previous run to pre-populate page table entries (PTEs) for the next, thereby minimizing tracer-induced artifacts.

Our implementation includes a special network-backed file interface to ensure the restored function can begin useful work immediately. Used by per-language shims, this interface allows an invocation request to be queued and made ready before the restore completes. The host OS monitors thread interaction with this interface to identify critical request-handling threads and grants them the highest scheduling priority upon restore, ensuring they execute on their first available time slice.

To minimize snapshot size, we combine application-level cooperation with OS-level optimizations. The network-backed interface provides a channel for the OS to signal an impending checkpoint, allowing the application to proactively perform state trimming. This includes running garbage collection cycles and explicitly dropping pages belonging to freed caches or other transient data structures.

Separately, as a part of any snapshot operation, our OS implementation pursues its own optimizations to minimize unnecessary state. For example, it translates any MADV\_FREE calls (lazy memory freeing) into eager MADV\_DONTNEED calls to release memory immediately. Additionally, it trims each thread's stack by identifying the current stack pointer and discarding any data in the unused region above its redzone. These complementary techniques ensure the final snapshot is as lean as possible.

%% file: sections/tab-workloads.tex
\begin{table*}[ht]
        \centering
        \resizebox{\textwidth}{!}{\begin{tabular}{c c c || c c | c c | c c | c c | c c || l}
        \hline
        \multirow{2}{*}{\textbf{Language}} & \multirow{2}{*}{\textbf{Function}} & \textbf{Warm} & \multicolumn{10}{c||}{\textbf{Snapshot (Working Set)}} & \multirow{2}{*}{\textbf{Description}}\\ & &
        \textbf{Latency ($\mu$s)} & \multicolumn{2}{c}{\textbf{VMAs}} & \multicolumn{2}{c}{\textbf{Delta Intervals}} & \multicolumn{2}{c}{\textbf{Private Pages}} & \multicolumn{2}{c}{\textbf{Shared Pages}} & \textbf{Zero Pages}  & \textbf{Working Set (MB)} & \\
        \hline
        \hline
        \multirow{2}{*}{Java}    & \texttt{image}     & {255,713} & 258  & (118) & 708  & (253) & 78426 & (8378)  & 40,951  & (1997) & \multicolumn{1}{c|}{(5867)} & (63.4) & {Rotate a JPEG} \\
                                 & \texttt{mtml}                & 5658 & 189  & (54)  & 549  & (99)  & 6310  & (900)   & 40588  & (519)  & \multicolumn{1}{c|}{(3)}    & (5.6)  & {Matrix Multiplication} \\
        \hline
        \multirow{2}{*}{NodeJS}  & \texttt{image}     & 43,069 & 310  & (216) & 566  & (330) & 34166 & (7894)  & 27323  & (1708) & \multicolumn{1}{c|}{(4822)} & (56.3) & {Rotate a JPEG} \\
                                 & \texttt{json}      & 8376 & 290  & (163) & 469  & (200) & 9552  & (3655)  & 27323  & (1347) & \multicolumn{1}{c|}{(57)}   & (19.8) & {JSON (de)serialization} \\
        \hline
        \multirow{10}{*}{Python} & \texttt{html}             & 10,977 & 81   & (30)  & 193  & (87)  & 3806  & (1591)  & 3691   & (473)  & \multicolumn{1}{c|}{(64)}   & (8.3)  & {HTML rendering} \\
                                 & \texttt{cnn}          & {53,686} & 1912 & (176) & 3142 & (564) & 72804 & (10318) & 367651 & (3464) & \multicolumn{1}{c|}{(413)}  & (55.4) & {CNN inference} \\
                                 & \texttt{hello}            & 77 & 48   & (16)  & 154  & (56)  & 1105  & (168)   & 2405   & (176)  & \multicolumn{1}{c|}{(1)}    & (1.3)  & {A no-op function} \\
                                 & \texttt{image}     & 15,476 & 192  & (47)  & 328  & (97)  & 5169  & (1471)  & 5343   & (539)  & \multicolumn{1}{c|}{(10)}   & (7.9)  & {Rotate a JPEG} \\
                                 & \texttt{json}      & 7287 & 154  & (40)  & 272  & (93)  & 5002  & (1606)  & 3999   & (492)  & \multicolumn{1}{c|}{(22)}   & (8.3)  & {JSON (de)serialization} \\
                                 & \texttt{lr}           & 1095 & 1292 & (98)  & 1751 & (170) & 25207 & (1928)  & 39852  & (578)  & \multicolumn{1}{c|}{(13)}   & (9.8)  & {LR inference} \\
                                 & \texttt{pyaes}                 & 1667 & 49   & (16)  & 157  & (50)  & 1240  & (302)   & 2401   & (211)  & \multicolumn{1}{c|}{(1)}    & (2.0)  & {AES encryption} \\
                                 & \texttt{rnn}          & 8634 & 312  & (57)  & 1437 & (164) & 52382 & (1537)  & 762571 & (1013) & \multicolumn{1}{c|}{(49)}   & (10.2) & {RNN inference} \\
                                 & \texttt{video} & 156,751 & 399  & (83)  & 726  & (191) & 7966  & (1990)  & 56421  & (1382) & \multicolumn{1}{c|}{(1137)} & (17.6) & {Grayscale conversion} \\
        \hline
        \end{tabular}}
        \caption{Characterization of memory usage for various serverless functions. Delta intervals counts the number of contiguous sets of pages with modified/private application data. Values in parentheses refer to the working set, as opposed to the whole of the snapshot. For zero pages, only those in the working set are reported.
        }
        \label{tab:workloads}
\end{table*}

%% file: sections/eval.tex
\section{Evaluation}
\label{sec:eval}

\begin{figure*}[ht]
    \centering
    \includegraphics[width=\linewidth]{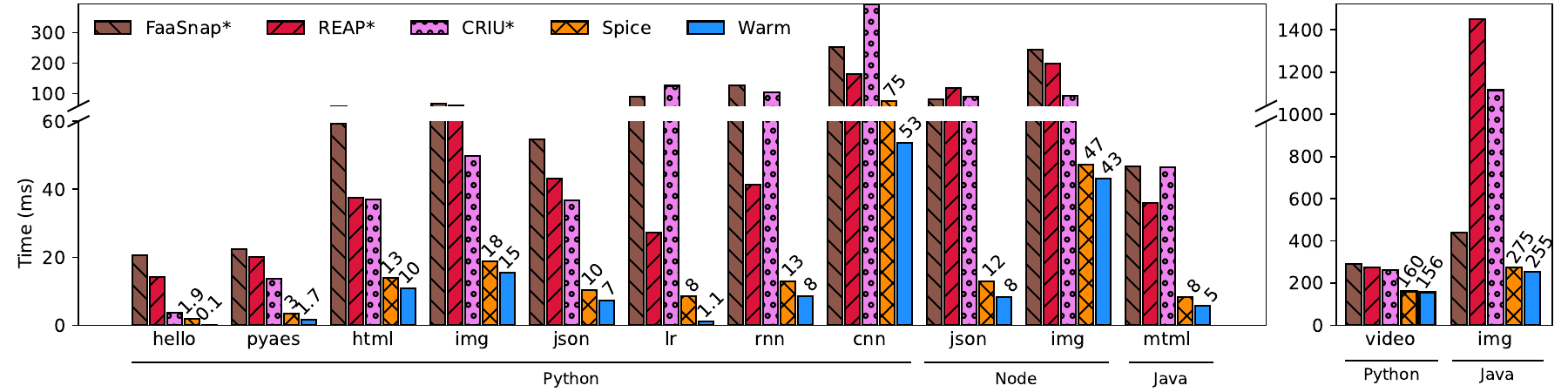}
    \caption{
    \sysname achieves end-to-end cold start latencies significantly closer to warm invocations than existing systems.
}\label{fig:e2e}
\end{figure*}

We aim to evaluate \sysname by answering the following questions:

\begin{enumerate}
    \item How does \sysname reduce end-to-end cold start latency in the context of existing snapshot/restore systems (\S\ref{subsec:e2e})?
    \item How does each technique introduced by \sysname contribute to reducing cold start latency (\S\ref{subsec:microbenchmarks})?
    \item How does \sysname perform in response to bursts of invocations (\S\ref{subsec:concurrency})?
\end{enumerate}

\paragraph{Experimental Setup} All experiments are conducted on a machine with an Intel(R) Xeon(R) Gold 5420+ with 28 cores @ 2.00GHz and 128 GB of RAM. Our storage device is a Crucial T705 NVMe drive with a max sequential read bandwidth of 13,600MB/s over PCIe Gen 5.0. We run all experiments from this drive and use it to store all checkpoint images and shared libraries used during function invocations. Our test suite includes the memory and CPU intensive functions from FunctionBench~\cite{functionbench}. FunctionBench is originally written in Python so we ported two~\nextdraft{\ben{add more if time?}} functions each to Node.js and Java to better capture the landscape of serverless functions. Table~\ref{tab:workloads} summarizes the functions used to evaluate~\sysname.

\subsection{End-to-End Latency}
\label{subsec:e2e}
To understand how \sysname performs in the context of existing systems, we compare to existing snapshot/restore systems that restore checkpoints entirely from storage without relying on any warm state. As discussed in Section~\ref{sec:background}, these systems include VM-based systems that introduce working set prefetching, FaaSnap~\cite{faasnap} and REAP~\cite{reap}, as well as CRIU~\cite{criu} which restores processes entirely from userspace. In light of the observations made in Section~\ref{sec:background} we augment the daemon responsible for executing functions in VM-based systems to run in the \texttt{SCHED\_FIFO} scheduling class to minimize interference from other tasks running the guest and call them FaaSnap* and REAP*. Native CRIU does not implement working set estimation but instead eagerly installs all memory pages. We modified CRIU to instead use \texttt{mmap} and restore memory with demand paging which lowers total latency compared to fetching all of checkpointed memory. We refer to our modified version of CRIU as CRIU*.
\nextdraft{\ben{should we include our userspace-only baseline with prefetching and explain that it shows the best we can do from userspace with respect to memory restore but includes our metadata optimizations? that might quell reviewers comments about unfairly comparing to CRIU}.}

Figure~\ref{fig:e2e} shows end-to-end function invocation latency using checkpoints restored entirely from storage with a cold page cache. \sysname is able to reduce latency significantly in all cases, by 4-89\% compared to REAP* and 7-92\% compared to CRIU*. We additionally compare \sysname to a warm function invocation. A warm invocation is one of a function that has already been invoked several times but with cold micro-architectural state (CPU caches, TLB). Cold caches illustrate an invocation on an otherwise busy system where cache state has been polluted by other processes running on the machine. With the exception of \texttt{hello} which does no computation, \sysname is 1.01-7.75$\times$ slower than a warm invocation; much of the added cost is due to VMA creation which cannot be parallelized with execution.

\sysname is particularly impactful for functions with short execution times, which represent the majority of functions in many serverless deployments~\cite{afaas, azure-traces}, but also reduces latency for functions with long execution times. This is a consequence of introducing a dedicated kernel interface -- functions with short execution times benefit from optimizations with smaller absolute impact on restore latency, like system call batching and Overlay VMAs, while functions with long execution times benefit from pre-installing PTEs and the page pool. For example, video processing has a large number of anonymous zero pages in its working set and can quickly retrieve a batch of free pages from the page pool for which the prefetcher thread will install PTEs. 

\begin{figure}[t]
    \centering
    \includegraphics[width=\linewidth]{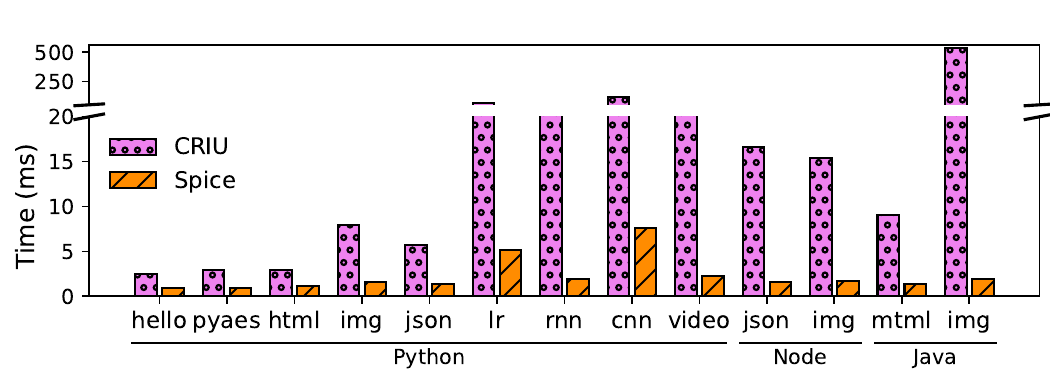}
    \caption{OS Metadata restore time comparing CRIU to \sysname. \sysname lowers latency through a dedicated interface for restoring OS state and compact serialization.}\label{fig:metadata}
\end{figure}

    

\subsection{Microbenchmarks}
\label{subsec:microbenchmarks}

\textbf{Metadata Restore.} 
A key component of \sysname is the addition of a dedicated interface for restoring OS metadata including threads, VMAs, file descriptor state, signal handlers, and timers. \sysname implements a streamlined metadata restore step that leverages the Junction libOS to demonstrate restoration of these components without expensive the system call replay performed in CRIU. 
In CRIU, metadata restore consists of replaying the system calls that executed during a cold start and expensive de-serialization as user-space data structures are de-serialized and injected into the kernel through system calls and translated into kernel data structures.  In \sysname, the majority of metadata is held in the userspace data structures of the Junction libOS, while VMAs are recreated in bulk through the kernel module. 

Figure~\ref{fig:metadata} compares the metadata restore latency in CRIU to that of \sysname, including the time to create VMAs. In all cases, \sysname's metadata restore latency is significantly lower than that of CRIU, helping \sysname greatly reduce restore latency.

\begin{figure}[t]
    \centering
    \includegraphics[width=\linewidth]{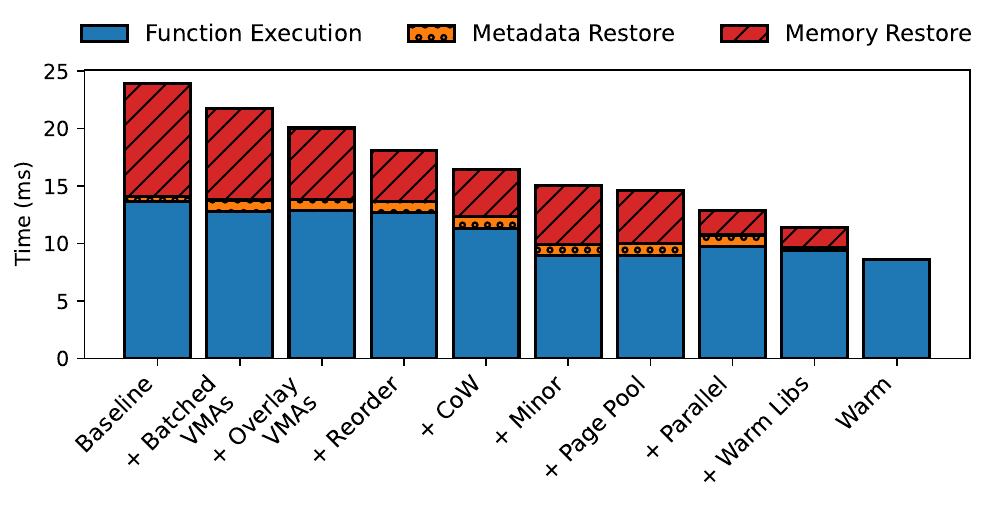}
    \caption{Ablation study of \sysname's memory restore optimizations on the RNN serving Python function.}\label{fig:ablation}
\end{figure}

\textbf{Memory Restore.} 
To evaluate \sysname's improvements to restoring memory, including actual memory contents, we measure a cold restore through our kernel module with all optimizations disabled and incrementally enable optimizations to illustrate their impact. Figure~\ref{fig:ablation} shows the results with the RNN serving function which performs inference on a small model. Each optimization contributes to an overall large reduction in memory restore time. Overlay VMAs reduce the number of VMAs that need to be created on restore when restoring mappings from shared files that become fragmented when written to. For RNN serving, 1451 VMAs would need to be created to overlay private/modified pages over a shared mapping. The original process has only 314 VMAs in total. While batched VMA creation reduces the impact of creating a large number of VMAs, Overlay VMAs contribute to reducing memory restore time. 

We see additional benefit from reordering the private working set pages in the checkpoint image to be placed in the order they will be accessed by the restored function. Reordering introduces additional VMA fragmentation because adjacent page ranges in the checkpoint file that would be mapped contiguously into a single VMA are split due to accesses that are temporally adjacent but not spatially adjacent. Reordered regions need to be re-shuffled to be adjacent again in the address space, which would necessitate one VMA per region. Reordering in this case creates 3095 total intervals, which we avoid creating through our use of Overlay VMAs.

\begin{figure}[t]
    \centering
    \includegraphics[width=\linewidth]{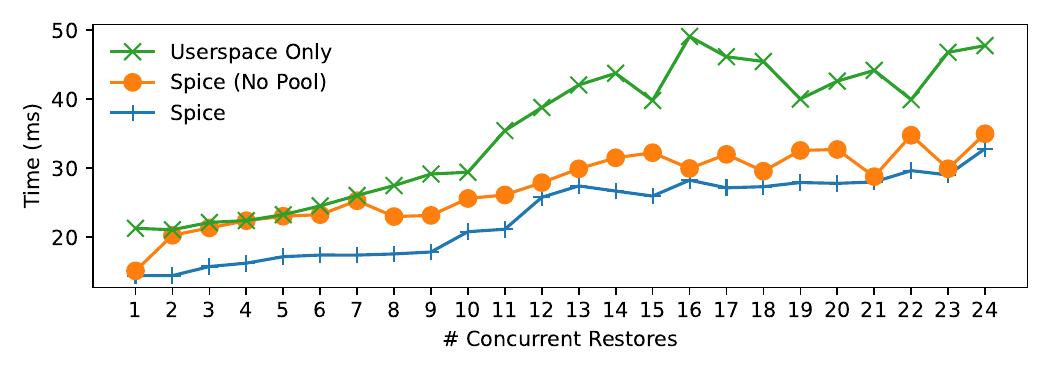}
    \caption{Maximum invocation latency as the number of concurrent invocations increases. \sysname maintains lower latency compared to an optimized userspace-only baseline due to careful kernel optimizations and lower contention.}\label{fig:parallel_restore}
\end{figure}
\subsection{Concurrency}
\label{subsec:concurrency}

To see how the techniques introduced by \sysname scale with bursts of invocations, we measure the maximum invocation latency 
in a burst of invocations of a single function from a cold page cache in three configurations: a userspace-only baseline that uses our metadata restore interface but restores memory entirely in userspace, \sysname (no pool) uses our memory restore interface with the page pool disabled so private pages are prefetched through the page cache, and \sysname with all optimizations enabled. We find that \sysname is able to maintain significantly lower latency as the number of concurrent invocations increases compared to userspace. We also find that the addition of the page pool further reduces latency even though it issues more I/O to read private pages that will be written to during restore. This is likely due to avoiding large page allocations on the critical path, leading to lower latency and more stable performance.

%% file: sections/related.tex
\section{Related Work}\label{sec:related}

\textbf{Remote fork.} Remote fork techniques for function cloning have been
explored recently~\cite{mitosis,cxlfork,trenv}. These systems assume a function
is warm in the datacenter and use hardware innovations like RDMA and
CXL to quickly fork a new instance on a different machine. We consider these
approaches to be orthogonal to ours: in the case that no root function is warm
in the datacenter, restoring it from disk is preferable to starting a fresh new
instance.

\textbf{Sandboxing.} Existing systems use lightweight VMs \cite{reap, sabre,
firecracker, faascache, faasnap}, CRIU \cite{replayable-exec, prebaking,
pronghorn}, or containers \cite{pagurus, medes} for checkpoint/restore. As we
discussed in Section \ref{sec:background}, these approaches suffer from
performance challenges. Some systems use alternative approaches for sandboxing.
Faasm \cite{faasm} relies on Wasm runtime as the isolation mechanism, which
offers low startup costs but higher end-to-end execution time compared to a
native Linux execution due to the cost of SFI \cite{wasm-slow}. SEUSS
\cite{seuss} implements unikernels, tailoring their sandbox to executing a
specific function. Unlike \sysname, SEUSS requires backporting to support
additional language runtimes, complicating deployment on existing FaaS
platforms.

\textbf{Prefetching.} Prefetching the working set of
a function to reduce execution time has been explored by prior work~\cite{sabre, reap, faascache}. Because there
has yet to be a sufficient OS interface for restoring and prefetching function
state, these systems suffer from additional overhead with the use of existing
solutions. \sysname demonstrates that these overheads are not fundamental and
other systems could benefit from our design, for example, by snapshotting and
restoring a function running in \sysname inside a VM if virtualization is
desired. 


\textbf{Avoiding cold starts.} Other systems suggest approaches for pre-warming, optimizing keep-alive policies, and container re-use to avoid cold starts \cite{rlpredict,
icebreaker, xanadu, faascache, keep-alive, jvmwarm, rainbowcake, orion, pagurus}. Keeping sandboxes warm
hurts resource elasticity, and \sysname's fast cold starts represent a step toward eliminating the need for keep-alive policies.
Other systems optimize the snapshot timing to ensure that the captured state is the function at its highest performance~\cite{seuss,serverless-java-runtime}.
Faascale~\cite{faascale} and AFaas~\cite{afaas} both discuss costs associatd with VM EPT faults and pursue strategies to mitigate those. 

\textbf{Memory Deduplication.} Avoiding memory duplication is key for both performance and resource utilization. Fork-based approaches naturally share memory by leveraging CoW semantics~\cite{sock, sand, catalyzer, mitosis, cxlfork, trenv}. Other systems like SEUSS~\cite{seuss}, Medes~\cite{medes}, and AFaaS~\cite{afaas} propose explicit strategies for de-duplicating snapshot state. SEUSS and AFaaS create layered snapshot stacks to reuse shared components like language runtimes. However, because these stacks require a strict lineage of changes, their mechanisms are practically confined to sharing a single common base, but not the complex, overlapping mixtures of libraries found across different functions. The Medes system~\cite{medes} takes a different approach; it reduces the memory penalty of keep-alive containers using explicit deduplication to allow more functions to remain in a semi-warm state. In contrast, \sysname aims to reduce the need for keep-alive policies altogether by enabling function state to be stored to and rapidly restored from disk.

\textbf{Control path optimization.} While \sysname focuses on the data path of function instantiation, significant overheads can also arise from the control path, including request scheduling, resource placement, and network setup. Systems like Dirigent, AFaaS, SigmaOS, and others have explored sophisticated schedulers and resource managers to minimize these orchestration latencies~\cite{dirigent, afaas, sigmaos, faasnet, atoll, flashcube}. These efforts are orthogonal and complementary to our work. Achieving the lowest possible end-to-end latency requires a holistic approach, combining an efficient control path with the rapid instance restoration provided by \sysname.

%% file: sections/discussion.tex
\section{Discussion}
Our evaluation demonstrates that by co-designing a snapshot/restore system with new OS primitives, \sysname can reduce cold start latency to under 5ms. These results challenge the conventional wisdom that restoring from persistent storage is fundamentally slow. This opens up new possibilities for the design of serverless platforms and raises important questions for future work.

\textbf{Resource Management and Keep-Alive Policies.}
\sysname's sub-5ms restore times in our prototype blur the line between warm and cold starts, potentially altering the economics of serverless resource management. The primary motivation for expensive keep-alive pools—avoiding the high latency of a cold start—is significantly diminished. This enables a new operational model where platforms can practice aggressive reclamation of idle instances to boost utilization, relying on just-in-time instantiation from disk to meet traffic demands without a major latency penalty.

This model also unlocks new strategies for cluster-level optimization. Because \sysname leverages the host page cache for sharing, operators can create specialized node pools dedicated to functions with similar software stacks (e.g., a ``Python+AI'' pool). Co-locating these functions maximizes the page cache hit rate for common runtimes and libraries, which both accelerates restores and increases overall memory density through natural deduplication.

\textbf{Integration with Fork-Based Approaches.}
The snapshot/restore model of \sysname is complementary to the cloning model of fork-based scaling mechanisms~\cite{mitosis, cxlfork,catalyzer}. Whereas \sysname is optimized for rapidly instantiating the initial ``parent'' process from persistent storage, fork-based systems excel at cloning that warm parent at microsecond latencies for horizontal scaling.

This relationship suggests a new, hybrid architecture for function instantiation. The principles of \sysname's page-cache-aware design could be extended across the network to create a distributed page cache. Such a service would maintain a rack- or cluster-wide pool of frequently-accessed memory pages from common libraries and runtimes. During a restore, \sysname could then source required pages from this low-latency remote memory fabric in addition to local storage, further reducing instantiation times and improving resource utilization across the cluster.

\textbf{Limitations and Future Work.}
A key next step is extending our approach to virtualized environments. While our current techniques can already improve restore times inside a guest VM by reducing in-guest overheads, further potential can be unlocked with direct host-guest cooperation. This could be achieved with a custom hypercall interface that allows the guest to eagerly request EPT population for a dispersed working set. By enabling the guest to pass its predicted memory layout to the hypervisor in a single, batched operation, this approach would eliminate the thousands of costly VM exits typically caused by individual page faults during memory restoration.

Furthermore, while we evaluated with fast local storage, our techniques could be adapted to operate over the network. This would enable restoring functions from remote storage or directly from another node's memory, blurring the line between cold starts and remote fork systems and offering greater placement flexibility in large-scale clusters.

%% file: sections/conclusion.tex
\section{Conclusion}
By demonstrating that cold starts from persistent storage can achieve near-warm latency, this work redefines the fundamental trade-offs in serverless computing. The long-accepted compromise between performance and memory elasticity is not an inherent limitation, but an artifact of operating systems designed for an era before serverless. Our findings suggest that the focus of optimization should shift from user-space heuristics and keep-alive policies to first-class OS support for state restoration.

\sysname serves as a blueprint for this new direction. By co-designing the execution engine with the kernel, we unlock a new operational model where functions can be aggressively offloaded to disk to maximize density and efficiency, yet instantiated in milliseconds on demand. This approach opens avenues for future platform architectures built around just-in-time, disk-based instantiation as the default, rather than the exception.